\documentstyle[12pt,aaspp4,epsfig]{article}

\begin{document}

\def \cmm  {cm$^{-2}$}
\def \cmmm {cm$^{-3}$}
\def \kms  {km~s$^{-1}$}
\def \Lya  {Ly$\alpha$}
\def \lya  {Ly$\alpha$}
\def \Lyb  {Ly$\beta$}
\def \Lyg  {Ly$\beta$}
\def \nhi  {$N_{HI}$}
\def \nhe  {$N_{HeII}$}
\def \object {Q0130-4021}

\def \hethree {$^3$He}
\def \hefour  {$^4$He}
\def \liseven {$^7$Li}

\title{QSO 0130-4021: A third QSO showing a low Deuterium to Hydrogen Abundance Ratio}

\author{
	David Kirkman\altaffilmark{1,2},
	David Tytler\altaffilmark{1,3}, \\
        Scott Burles\altaffilmark{1,4} \\
	Dan Lubin\altaffilmark{1} \& John M. O'Meara\altaffilmark{1} \\
Center for Astrophysics and Space Sciences;
\\ University of California, San Diego; \\ MS 0424; La Jolla; CA
92093-0424\\}
 
\altaffiltext{1} {Visiting Astronomer, W.M. Keck Observatory which
is a joint facility of the University of California, the California
Institute of Technology and NASA.}
\altaffiltext{2} {E-mail: dkirkman@physics.bell-labs.com, present address: 
Bell Laboratories, Lucent Technologies, 600 Mountain Avenue, Murray Hill, NJ 07974-0636}
\altaffiltext{3} {E-mail: tytler@ucsd.edu}
\altaffiltext{4} {Present address: Univ. of Chicago, Astronomy \& Astrophysics
Center, 5460 S. Ellis Ave., Chicago, IL 60615}

\abstract

We have discovered a third quasar absorption system which is
consistent with a low deuterium to hydrogen abundance ratio, D/H $=3.4
\times 10^{-5}$.  The $z_{abs} \sim 2.8$ partial Lyman
limit system towards \object\ provides the strongest evidence to date
against large D/H ratios because the H I absorption, which consists of
a single high column density component with unsaturated high order Lyman
series lines, is readily modeled -- a task which is more complex in other 
D/H systems. We have obtained twenty-two hours of spectra from the HIRES
spectrograph on the W.M. Keck telescope, which allow a detailed
description of the Hydrogen.  We see excess absorption on the blue
wing of the H I \lya\ line, near the expected position of deuterium.
However, we find that Deuterium cannot explain all of the excess
absorption, and hence there must be contamination by additional
absorption, probably H~I. This extra H~I can account for most or all
of the absorption at the D position, and hence D/H $=0$ is allowed.
We find an upper limit of D/H $\leq 6.7 \times 10^{-5}$ in this system,
consistent with the value of D/H $\simeq 3.4 \times 10^{-5}$ deduced 
towards QSO 1009+2956 and QSO 1937-1009 by Burles and Tytler (1998a,
1998b).  This absorption system shows only weak metal line absorption,
and we estimate [Si/H] $\le -2.6$ -- indicating that the D/H ratio of the
system is likely primordial.  All four of the known high redshift
absorption line systems simple enough to provide useful limits on D
are consistent with D/H = $3.4 \pm 0.25 \times 10^{-5}$. Conversely,
this QSO provides the third case which is inconsistent with much
larger values.

\section{Introduction}

Big bang nucleosynthesis (BBN) is a cornerstone of modern
cosmology. The standard theory (SBBN) predicts the abundances of the
light nuclei (H, D, \hethree, \hefour, and \liseven) as a function of
a single parameter, the cosmological baryon to photon ratio, $\eta
\equiv n_b/n_\gamma$ (Kolb \& Turner 1990).  The ratio of any two
primordial abundances should yield a direct measurement of $\eta$, and
the measurement of a third provides a test of the theory.

The abundances of all three of the light elements have been measured
in a number of terrestrial and astrophysical environments, however,
most of these are not primordial abundances, and corrections for later
chemical evolution are problematic.  Adams (1976) suggested that it
might be possible to measure primordial D/H towards low metallicity
absorption line systems in the spectra of high redshift QSOs.  The
task proved too difficult for 4-m class telescopes (Chaffee et
al. 1985, 1986; Carswell et al. 1994), but became possible with the
advent of the HIRES echelle spectrograph (Vogt 1994) on the W.M. Keck
telescope.  Using Keck+HIRES, Songaila et al. (1994) reported an upper
limit of D/H $< 25 \times 10^{-5}$ in the $z_{abs} = 3.32$ Lyman limit
system (LLS) towards QSO 0014+813.  Using different spectra, Carswell
et al. (1994) reported $<60 \times 10^{-5}$ in the same object, and
they found no reason to think that the deuterium abundance might be as
high as their limit.  Chaffee et al. (1985, 1986) had reached the same
conclusion a decade earlier.  The most recent, improved spectra
(Burles et al. 1999) support the early conclusions: D/H $< 35 \times
10^{-5}$ for this QSO.

There are currently two known absorption systems in which D/H is low:
D/H = $4.0 \pm ^{+0.8}_{-0.6} \times 10^{-5}$ in the $z_{abs} = 2.504$ LLS
towards QSO 1009+2956 (Burles \& Tytler 1998a), and D/H $= 3.24 \pm
0.3 \times 10^{-5}$ in the $z_{abs} = 3.572$ LLS towards QSO 1937-1009
(Tytler, Fan, and Burles 1996; Burles \& Tytler 1998b).

There remains uncertainty over a fourth LLS at $z_{abs} = 0.701$
towards QSO 1718+4807, because we lack spectra of the Lyman series
lines which are needed to determine the velocity distribution of the
Hydrogen.  Webb et al. (1997a, 1997b) assumed a single hydrogen
component and found D/H = $25 \pm 5 \times 10^{-5}$. Levshakov et
al. (1998) allow for non-Gaussian velocities and find D/H $\sim 4.4
\times 10^{-5}$, while Tytler et al. (1999) find $8 \times 10^{-5} <$
D/H $<57 \times 10^{-5}$ (95\%) for a single Gaussian component, or
D/H as low as zero, if there are two hydrogen components, which is not
unlikely.

Very few LLS have a velocity structure simple enough to show
deuterium.  This can be understood if the LLS arise in forming
protogalaxies with potential wells of $\sim 200$ \kms, as expected in
most CDM hierarchical cosmologies (Cen and Ostriker 1993; Rauch et
al. 1997).  Since a merging protogalaxy is unlikely to make a simple
absorption system, we usually see one or two absorbers
with column densities $N_{HI} > 10^{16}$
\cmm\ and several with $N_{HI} > 10^{15}$ \cmm\ distributed 
over 200-300 \kms. These abosrbers usually absorb most of the QSO flux near
--82 \kms\ from \lya, where the D~I line is expected, and hence we obtain no
information of the column density of D~I.

In this paper, we present a third absorption system which
unambiguously shows low D/H.  The $z_{abs} = 2.799$ partial LLS
towards \object\ is simpler than the others because the entire Lyman
series is well fit by a single velocity component.  The velocity of
this component and its column density are well determined because many
of its Lyman lines are unsaturated. Its \lya\ line is simple and
symmetric, and can be fit, using zero free parameters, using only the
H~I parameters determined by the other Lyman series lines.  There is
barely enough absorption at the expected position of D to allow low
values of D/H, and there appears to be no possibility of high D/H.

\section{Observations and data reduction}

Our observations of QSO Tololo 0130--403 (Osmer \& Smith 1976;
redshift 3.023, V=17.02, B1950 RA 1h 30m 50.3s, Dec --40d 21m 51s;
J2000 1h 33m 1.96s --40d 6m 28s) with the HIRES echelle spectrograph
on the W.M. Keck-I telescope are described in Table \ref{obs}.  We observed
for 5 hours in 1995 and 1996 with the HIRES red cross disperser.  This
gave complete coverage between 3900 and 5200 \AA, continuing 
from 5200 -- 6000 \AA\ with small ($\sim 1-5$ \AA) gaps every 
40 -- 50 \AA\ between the spectral orders.
We also observed for another 17.3 hours
with the HIRES ultraviolet cross disperser, and obtained complete
spectral coverage between 3350 and 4850 \AA.  All spectra used the
HIRES C5 decker, which provides an entrance aperture to the
spectrograph with dimensions 7.5$^{''}$x1.15$^{''}$ and a spectral
resolution of 8 \kms.

All observations were made at telescope elevations of less than 30$^0$
because of the low declination of the QSO. 
The UV observations were made with the
paralactic angle aligned parallel to the spectrograph entrance slit
using the HIRES image rotator (Tytler et al. 1999), which was not not
available for the red observations.

The data were flat-fielded, optimally extracted, and wavelength
calibrated in the standard way using Tom Barlow's set of echelle
extraction (EE) programs.  Each observation was extracted and
wavelength calibrated separately.  The data were not flux calibrated,
because this cannot readily be done to high accuracy. 
Instead, as described
below, a local continuum was fit to each order of each observation.

\subsection{Wavelengths}

We performed several checks on the wavelength calibrations.  For each
spectral order, typically 20--30 Thorium or Argon reference lines were
identified and hence we expect wavelength errors of $<0.1$ pixels.
Several bright atmospheric OH emission lines were measured to lie within 
0.01 \AA\ (about 0.04 pixels) of their expected positions.

The HIRES cross-disperser has been found to drift by up to a few pixels,
especially when the telescope moves. Therefore, we obtained 
calibration spectra before and after each observation, but we found no
shifts when we cross-correlated these images.  We continuously monitored 
the position of the cross disperser, and we reset it when it moved, but 
we found no associated errors in wavelengths, in part because these 
shifts move spectra in the sky direction, along the length of the slit. 

Each observation was rebinned onto a wavelength scale that was linear
in velocity with 2.1 \kms\ pixels, and then combined into a single
spectrum.  Each pixel in the final spectrum is the weighted mean of
the pixels values in each of the individual observations.

\subsection{Continuum Placement}

Low order continuum fits were used to avoid the introduction of
artifacts.  Prior to summation, a local continuum was fit to each
order of each observation using the IRAF task CONTINUUM.  A third
order Legendre polynomial was fit to each order of each observation.
This is a lower order than is needed to fully describe the continuum
of a QSO over a HIRES echelle order.  Kirkman and Tytler (1997) found
that QSO continua regions with little absorption are typically
described by a Legendre polynomial of order 7 to 9 per HIRES order.
Here, we use third order polynomials for two reasons.  First, they
produce good continuum levels throughout most of the spectrum,
although there are some regions which may be off by $\sim$ 10
\%. Second, they give very similar continuum levels when applied to
different observations of the same object, and hence they should not
introduce artifacts when we combine the separate observations.

The IRAF CONTINUUM task was set to iterate until it found a stable
fit, each time rejecting pixels more than 1$\sigma$ below the fit,
and keeping all pixels above the fit.  This set of rejection
parameters results in the continuum task placing the continuum at the
top of each order.  The continuum task was
allowed to fit each order of each observation without any operator
supervision.  This resulted in obviously incorrect continuum estimates
in a number of locations throughout the each spectrum.  Looking at the
final combined spectrum, we checked the continuum level selected near
each absorption feature discussed in this paper.  With the exception
of the region near the $z_{abs} \sim 2.799$ H I Lyman limit, the
automatically selected continuum is the same we would have chosen by
eye, which is the standard way of continuum fitting QSO absorption
line spectra.  The procedure failed near the continuum discontinuity
of the H I Lyman limit, but as anticipated, it failed in the same way in
each of the individual observations and thus did not affect the shape
of any of the absorption features.  We hand corrected the continuum
level in this region.

\section{The $z_{abs} \sim 2.799$ partial Lyman Limit absorption system}

\object\ has a partial Lyman limit absorption system at $z_{abs} \sim
2.8$.  The lines associated with this system are tabulated in Table
\ref{dhlinetab}. This system is very simple, with only one major velocity
component.  We measure the H absorption and the D/H ratio first, and
then the metal abundance.

\subsection{Hydrogen and Deuterium absorption}

\begin{figure}
\centerline{\psfig{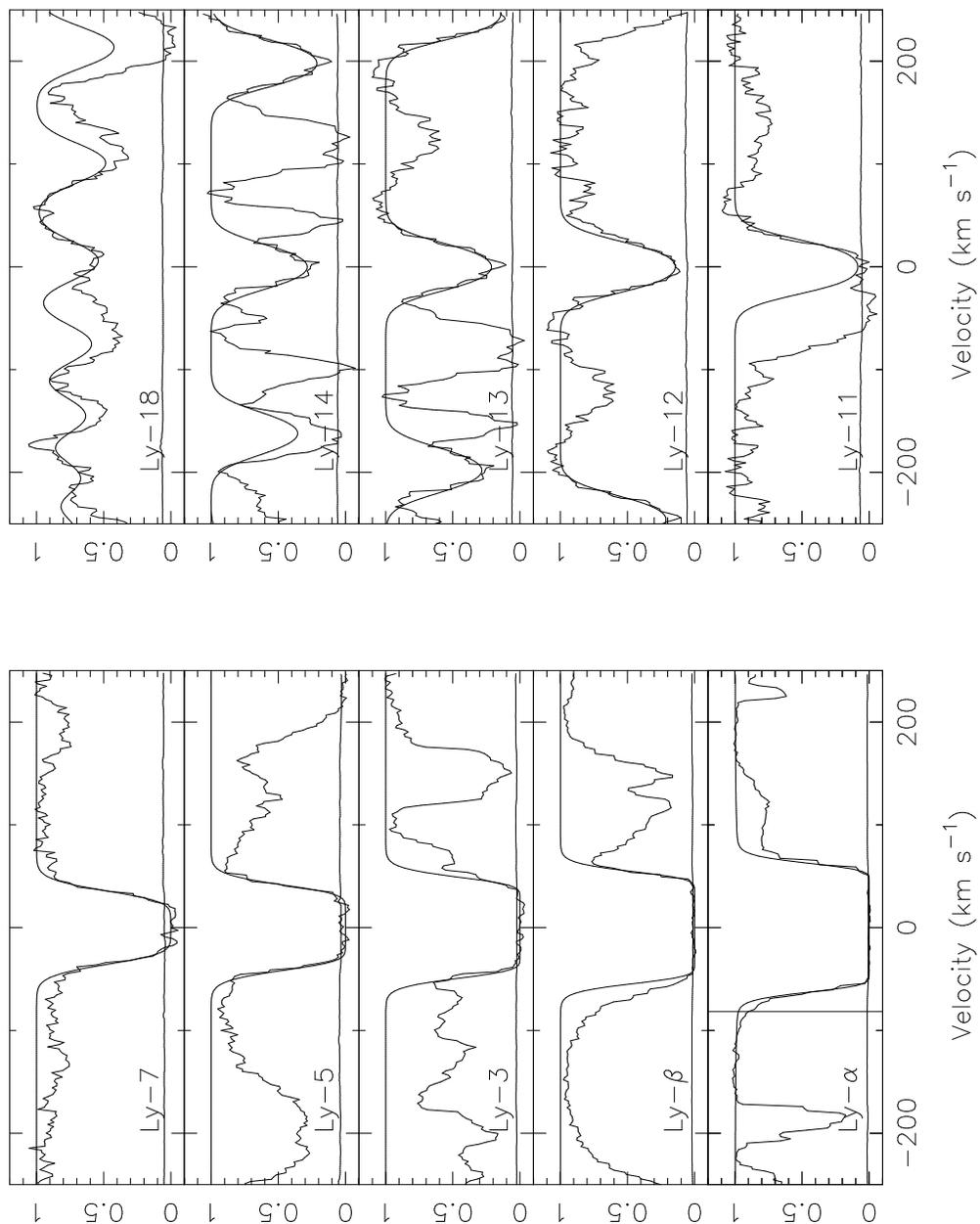}}
\caption{Lyman Series absorption in the $z_{abs} \sim 2.8$ absorber. The smooth
line is the fit to the H~I, using the parameters in Table \ref{dhlinetab}. 
The vertical line at --82 \kms on the \lya\ panel shows the expected
position of D. The line just above zero in these and other spectral plots
shows the 1$\sigma$ random error.}
\label{dhlyserfig}
\end{figure}

We can unambiguously determine the line parameters of the strongest
Hydrogen absorbers by looking at the unsaturated high order Lyman
series lines.  Absorption from the 
Lyman series is shown in Figure \ref{dhlyserfig}.  We have fit Voigt
profiles to Ly-11, Ly-12 and Ly-13 using the VPFIT code (Webb 1987) to
determine the H I absorption parameters of this system.  These lines
are all unsaturated and well fit by a single component at $z =
2.799742$. The column density, log \nhi $=16.66 \pm 0.02$ \cmm, is low, and
the line width parameter $b = 22.7\pm 0.1$ \kms\ is narrow for a LLS.
Figure \ref{dhlyserfig} shows that the other Lyman series lines
predicted by this fit are reasonable because they do not over
absorb. They frequently under absorb, which is not a problem, because
we do not try to fit the \lya\ forest absorption at other redshifts.

\begin{figure}
\centerline{\psfig{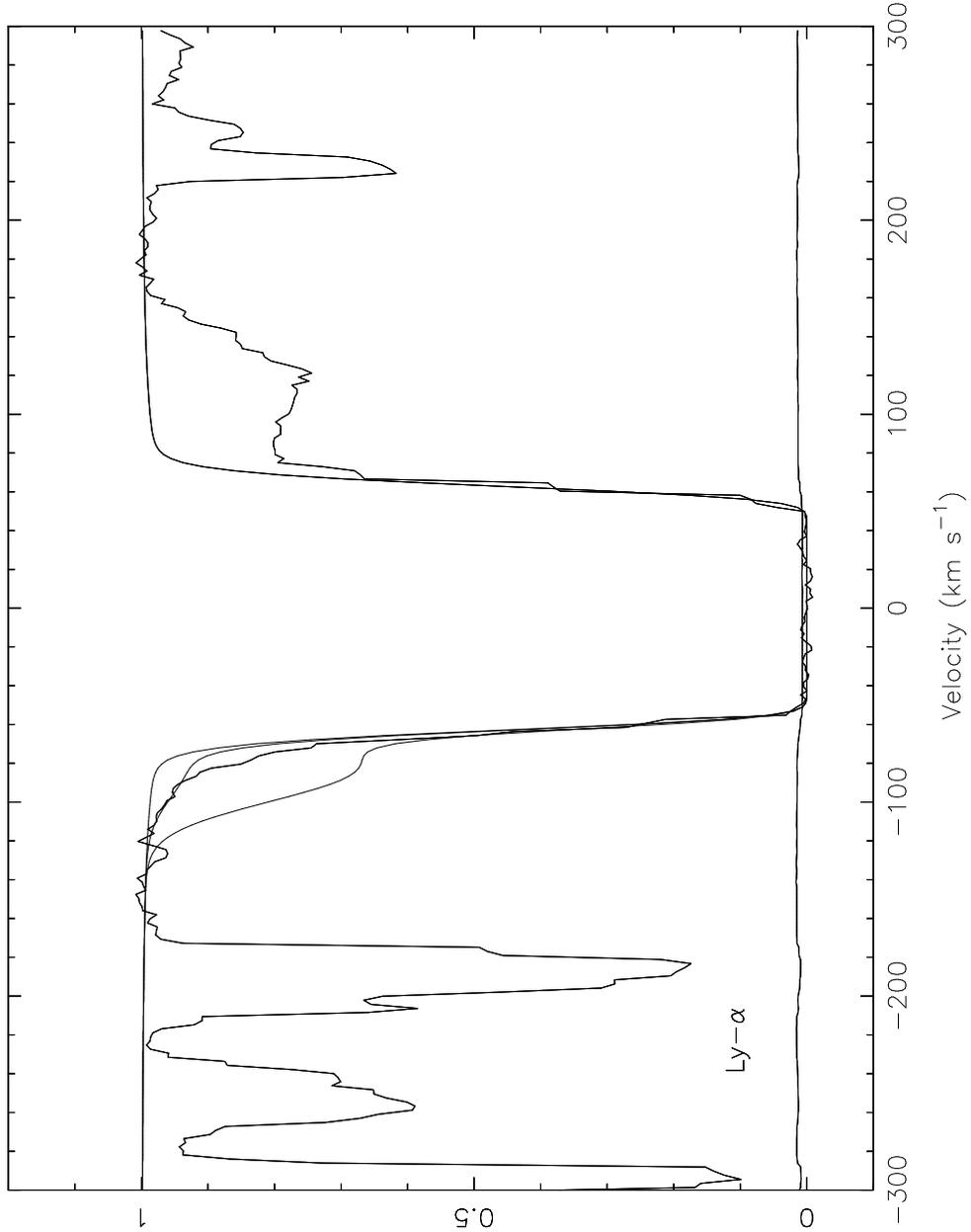}}
\caption{The \lya\ line of the $z_{abs} \sim 2.8$ LLS is centered at velocity 
$v=0$ \kms.  The vertical line at --82 \kms\ shows the expected
position of D.  Overlaid is the expected D+H absorption if D/H = 0
(upper), D/H = $3.4 \times 10^{-5}$ (middle), and D/H = $25 \times
10^{-5}$ (lower), assuming turbulent broadening.  The continuum
level is at unity on the vertical scale.}
\label{fdh0-3.4-25}
\end{figure}

Figure \ref{fdh0-3.4-25} shows the H I \lya\ line of the $z \sim 2.8$
absorption system, overlaid with the fit predicted from the Lyman
series lines Ly-11, Ly-12, and Ly-13, and three different values
for D/H: 0, $3.4 \times 10^{-5}$
and 25 $\times 10^{-5}$.  The saturated core and steep
sides of the \lya\ line are fit well, considering that no parameters
were adjusted to fit the \lya. 

The parameters of the D~I line are constrained by the H~I.  There
should be a single D~I component, because there is a single high
column density H~I component, and this D~I must be at the same
redshift as the H I. In the velocity frame of the H~I, the D~I line
will be centered at --82 \kms, or 4617.976 $\pm$ 0.004 \AA\ 
(heliocentric, vacuum value).

We can readily estimate D/H by comparing the amount of absorption
predicted with the spectra.  The D/H $=0$ under-absorbs near --82
\kms, and hence there must be additional absorption, probably H or D.
The D/H $=3.4 \times 10^{-5}$ prediction is consistent with the data. 
However, additional absorption is needed near --70 \kms, probably H~I centered
at --50 \kms.  Larger D/H does not help because it over-absorbs near
--82 \kms, as seen for D/H $= 25 \times 10^{-5}$.

\begin{figure}
\centerline{\psfig{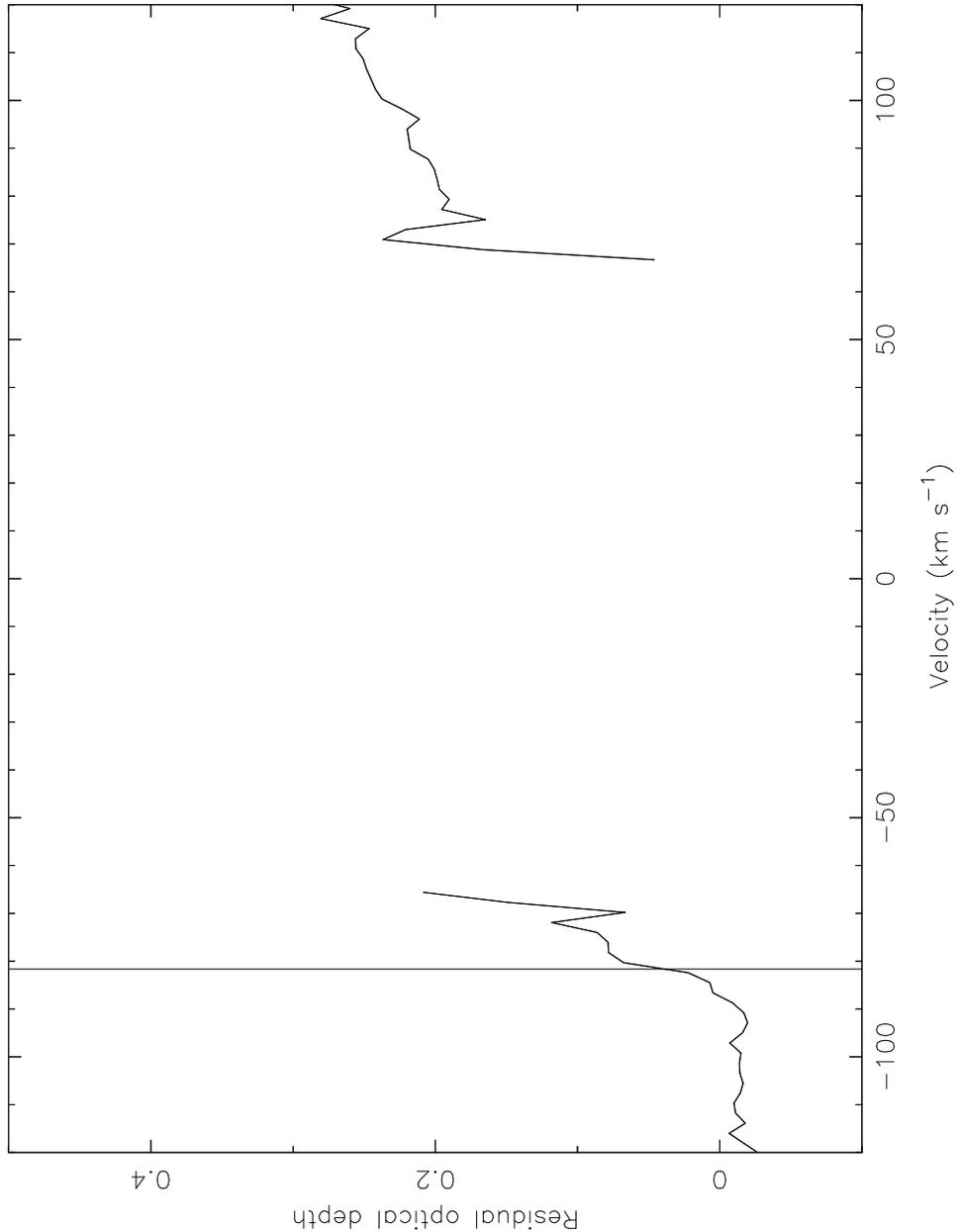}}
\caption{The excess absorption near the \lya\ line of the 
$z_{abs} \sim 2.8$ LLS.
We plot the residual optical depth between the observed data and the
H~I line predicted from the higher order Lyman lines, using the
parameters in Table \ref{dhlinetab}. The residual optical depth is not 
symmetric about --82 \kms, meaning that a D~I line cannot fully explain all of
the observed absorption on the blue wing of H~I \lya.  The residual
optical depth is poorly known, with large errors, near --65 \kms, and
is not known at $-50 < v < +50$ \kms\ where we observe zero flux.}
\label{dhresflxfig}
\end{figure}

We draw four conclusions from Figure \ref{fdh0-3.4-25}.  First, no
single line centered at --82 \kms\ can account for all of the excess
absorption from --100 to --60 \kms.  Second, since the two (or more)
lines are strongly blended it will be hard to partition the absorption
between them, and hence we do not expect to to make an accurate
measurement of D/H.  This is shown in Figure \ref{dhresflxfig} where
we see that residual optical depth after the removal of the main H is
not clearly separated into two lines.  Third, D/H $ < 3.4 \times
10^{-5}$ is acceptable, and fourth, large values of D/H are strongly
rejected.

\begin{figure}
\centerline{\psfig{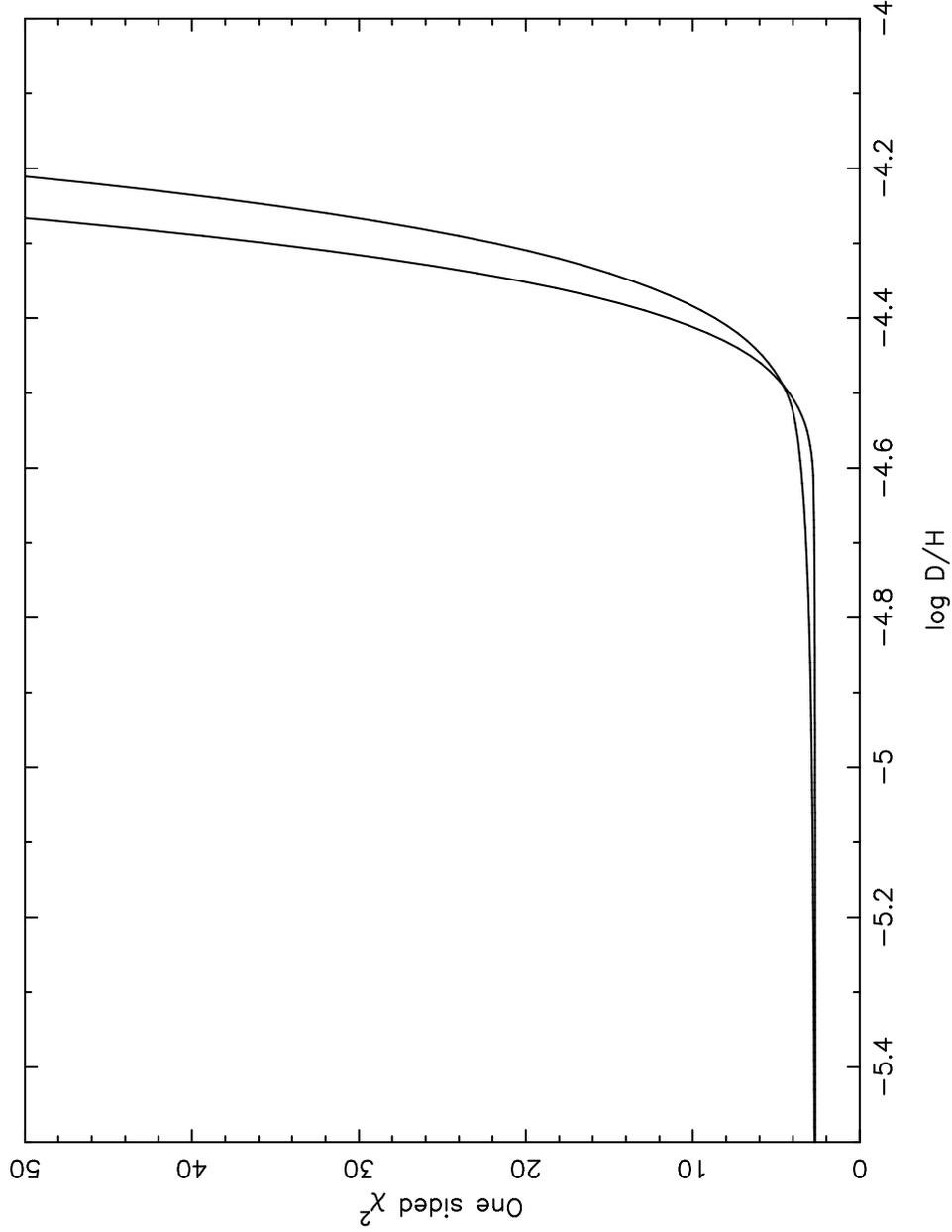}}
\caption{Goodness of fit of the model to the spectra 
in the region of the D line.  We plot $\chi^2$ vs D/H. The upper line
(at log D/H = --5) assumes turbulent line widths, while the lower line
assumes thermal line widths.}
\label{dhchiturbfig}
\end{figure}

We can place an interesting upper limit on the amount of D I present
in this system because there is little absorption near --82 \kms .  We
calculated a one sided $\chi^2$ statistic between the observed data
and the predicted H I + D I absorption as a function of D/H. For each
pixel, the contribution to the one sided $\chi^2$ is computed as a
normal $\chi^2$ statistic if the predicted flux lies below the data
(i.e. if there is too much absorption, which is unphysical, and hence
argues for rejection of the model) and is taken to be zero if the
prediction lies above the data (which could be a result of unrelated
absorption). We apply the statistic over the range 4617 - 4618.3 \AA\
(40 pixels), where the D line has the most effect on the fit. Since
the value of \nhi was set using the higher order Lyman series lines,
we can exclude wavelengths where H~I is the main absorption, because
the quality of the fit in this region will have little effect on D/H.
When D/H is zero, the fit lies above the data for nearly all
pixels. As D/H increases the fit drops, and more pixels are included
in the statistic.  We calculated this statistic twice, once assuming
$b_{D I} = b_{H I}$ and again assuming $b_{D I} = b_{H I}/\sqrt{2}$,
representing thermal broadening. The results are shown in Figure
\ref{dhchiturbfig}. The one
sided $\chi^2$ is essentially flat and constant for D/H $< 4.5 \times
10^{-5}$, and it increases rapidly with larger values of D/H.
A value of D/H $= 4.5 \times 10^{-5}$ is acceptable:
Prob( $\chi^2 > 20$ for 20 pixels) = 0.45.
The formal probabilities drop rapidly with increasing D/H, attaining
Prob($ \chi^2 > 48 $ for 21 pixels) = 0.0007 by D/H $< 5.4 \times 10^{-5}$.
We shall quote a limit of
D/H $< 4.8 \times 10^{-5}$ where the Prob($ \chi^2 > 29$ for 20 pixels) = 0.09.
This limit is from random errors alone, assuming the continuum level shown in 
Figure 1. 
The 9\% probability is an under-estimate because we ignored the 20 pixels
which lie below the fit for this D/H value. If we were to add an additional 
absorption line to the fit of this wavelength region, and we assume that 
each of the ignored pixels
then adds about unity to the $\chi ^2$ sum, then the 9\% would change to
about 18\%.

We now discuss the continuum level errors, which are probably the dominant
source of error since we are measuring a
shallow (5\% deep) D line in high signal to noise ratio spectra. There
are few pixels near the continuum within 300 \kms\ of the D feature,
which makes it difficult to estimate the error on the continuum level.
 From experience with similar spectra (Kirkman \& Tytler 1997), 
we suggest that the continuum level
error is about 2\%, but we do not know how to estimate the distribution of 
this error with the existing data. The continuum could be higher than we have
estimated, if the spectra near --150 \kms\ and +190 \kms\ are depressed
by weak shallow \lya\ absorption.  If it were raised by 2\%, there would be
approximately 40\% more optical depth at the position of the D line, and 
the upper limit on D/H would rise from $< 4.8 \times 10^{-5}$ to
$< 6.7 \times 10^{-5}$, which we consider to be similar to a $1 \sigma$ limit.

We cannot place a lower limit on the amount of D, because all the
absorption at the position of D could be H.  In the case of QSO
1009+2956 and QSO 1937-1009 (Tytler, Fan \& Burles 1996; Tytler \& 
Burles 1997; Burles \& Tytler 1998b), 
we could use line widths to show that we had detected
D. The absorption at the location of both those D lines was narrower
than any normal \lya\ forest line.  But in the present case the
absorption near D can be fit with $b \simeq 22$ \kms, which is a
common width for H in the \lya\ forest.

\subsection{Metal absorption}

\begin{figure}
\centerline{\psfig{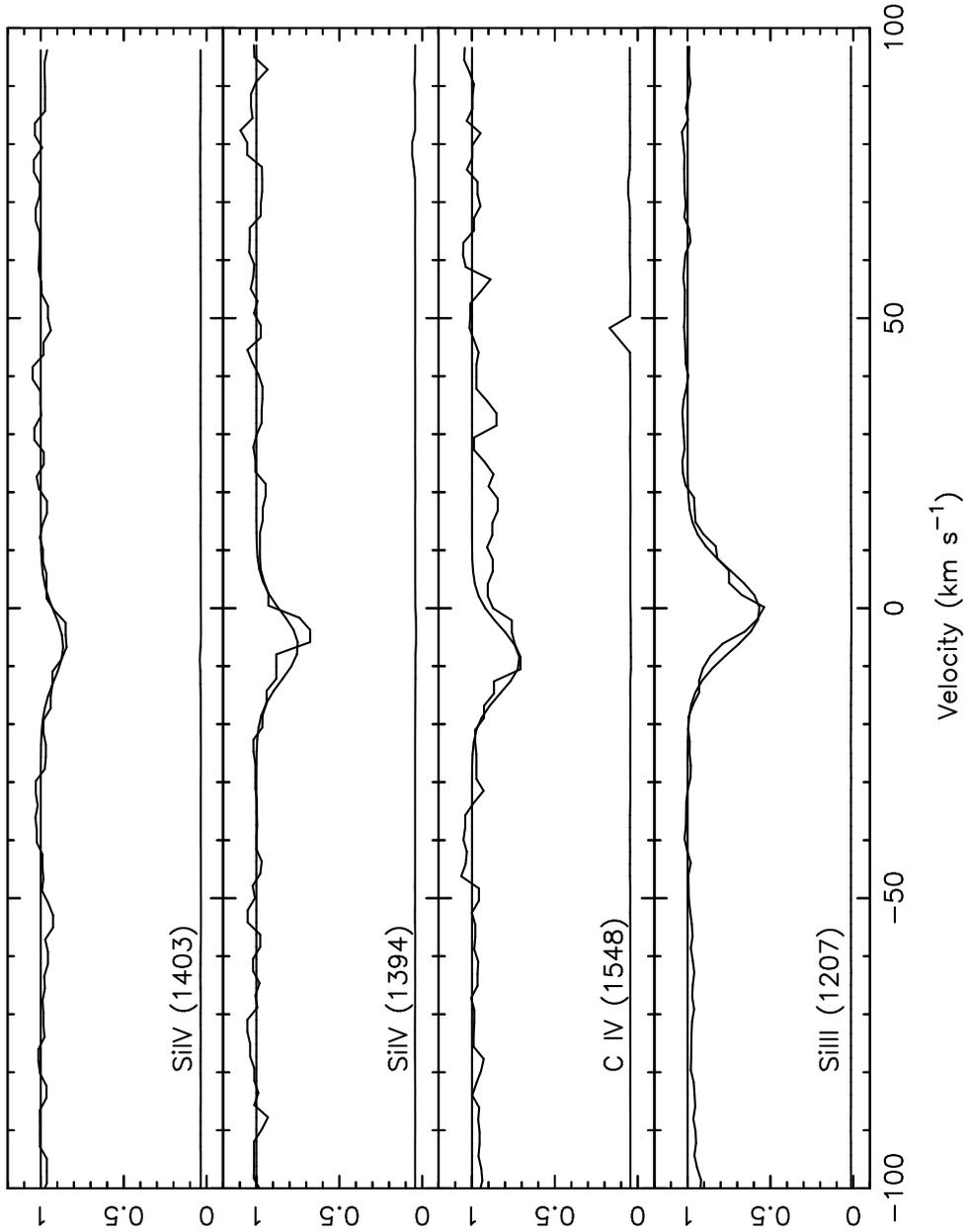}}
\caption{The metal absorption lines detected in the 
	$z_{abs} \sim 2.8$ absorber.  Overlaid are the Voigt profiles fit
	to the absorption, using the parameters in Table \ref{dhlinetab}.} 
\label{dhmetalfig}
\end{figure}

We detect metals at two velocities in this absorption system, Si~III
at $v \sim 0$, Si~IV and C IV at $v \sim -10$ \kms, as shown in Figure 
\ref{dhmetalfig}. The properties of
the detected ions are summarized in Table \ref{dhlinetab}.  The
Si~III line appears to be asymmetric, indicating that it may be a two
component blend.  However, we can only fit it as a single line because
our data are not sufficient to deblend the two components.  We did not
detect any C II, Si~II, C~III, N~V, O VI, or O I.  C~III could be
present, but we cannot detect it because that region of the spectrum
is strongly contaminated by unrelated \lya\ forest absorption.  The
lack of C II and Si~II is expected because the system is not optically
thick at the H I Lyman limit: -- C II and Si~II are generally 
found only in gas that is shielded from the metagalactic UV background
radiation below the H I Lyman limit.  N~V and O I are rarely observed,
so the lack of absorption in these species is not surprising either.

The H~I associated with the weak metal lines (C IV and Si IV) at $v \sim -10$
\kms, has a column density $N_{H I} < 10^{16}$ \cmm\ 
because it is not seen in the high order Lyman series lines,
especially Lyman-18, which are well fit with a single line at $v = 0$.
Thus the conclusions of the last section are unaffected by the second
velocity component at $v \sim -10$ \kms.

We estimate the metal abundance of the $v=0$ system by converting the
observed column density ratio Si~III/H~I into [Si/H].  If we make the
standard assumption that the absorber is in photoionization
equilibrium with the metagalactic UV background, we need to
determine the hydrogen density ($n_H$) of the absorber and specify the
spectrum of the UV background to make the conversion.

We get the $n_H$ from numerical simulations of structure formation in
CDM models.  These often find a tight correlation between the $N_{HI}$
of an absorption line, and $n_H$ in the gas producing the absorption.
Hellsten et al. (1997) found the relationship between column density
and physical density to be well approximated by
\begin{equation}
{\rm log} \; n_H = -14.8 + {\rm log} \; {\Omega_b h^2 \over 0.0125} 
	+ 0.7 \; {\rm log} \; N_{HI} 
\end{equation} 
in simulations of both SCDM and LCDM Universes.  Using $\Omega_b h^2 =
0.019$ and ${\rm log} \; N_{HI} = 16.66$ \cmm, we find that in this
absorber, the predicted density is ${\rm log} \; n_H = -2.9$ \cmmm.
Using CLOUDY (Ferland 1993)
with a Madau spectrum for the metagalactic UV background, the
observed [Si~III/H~I] implies [Si/H] = --2.6.  If [C/Si] $\sim 0.3$, as in
Population II stars, the solution is self consistent:  the predicted line
strengths of O~VI, C IV, Si~IV, C II and Si~II are low enough that they would 
not be detected in our data.  The overall low metallicity of the 
system implies that the gas has not been reprocessed, and thus
that our limit on D/H corresponds to gas which is primordial. 

\section{Conclusions}

To date, after the community has searched through more than 100 lines
of sight, only four had previously been found with simple enough
velocity structure to give useful limits on D/H.  All four of these
absorbers contain multiple strong H I velocity components.  Two of the
four yield a measurement of D/H, because they are free of major
contamination, while the other two can be strongly contaminated, and
allow D/H $=0$.

Here we presented the fifth absorption system which is simple enough
to show low D. The properties of the H are well determined. We find
that D/H $\simeq 3.4 \times 10^{-5}$ is acceptable, and that D/H
ratios $> 6.7 \times 10^{-5}$ are rejected.  This absorber provides the
strongest evidence yet against high D/H because it has unsaturated
lines and the simplest velocity structure, with only a single major
velocity component in H~I.  However, contamination by extra H
absorption near the D position remains a problem, because a small
column density of H~I could readily explain all of the absorption near
D. Hence D/H $=0$ is allowed.

The metallicity is apparently low, with [Si/H] $\simeq -2.6$, which
implies that the D has not been destroyed in stars. 

The limit D/H $< 6.7 \times 10^{-5}$ 
is in agreement with our two other measurements of the
primordial D/H: D/H $= 4.0^{+0.8}_{0.6} \times 10^{-5}$ towards
QSO 1009+2956 (Burles and Tytler 1998a) and D/H $= 3.24 \pm 0.3 \times
10^{-5}$ towards QSO 1937-1009 (Tytler, Fan, and Burles 1996; Burles
and Tytler 1998b).  We have argued that two other QSOs, QSO 0014+813
(Burles et al. 1999) and QSO 1718+4807 (Tytler et al. 1999) are also
consistent with this low D/H, and hence all QSO spectra are consistent
with a low primordial D/H ratio, D/H $\sim 3.4 \times 10^{-5}$.
However, QSOs 1009+2956 and 1937-1009 are inconsistent with D/H $\ge 5
\times 10^{-5}$, and now we have found that a third, \object, is
inconsistent with D/H $\ge 6.7 \times 10^{-5}$.

\acknowledgements                                   
This work was funded in part by grant G-NASA/NAG5-3237 and by NSF
grant 9420443.  We are grateful to Steve Vogt, the PI for the Keck
HIRES instrument, and to Theresa Chelminiak and Barbara Schaefer who
helped us obtain the Keck spectra.

\clearpage

\begin{deluxetable}{cccccc}
\tablecaption{\label{obs} Keck HIRES Spectra of \object\ }
\tablewidth{34pc}
\tablehead{
\colhead{Date} &
\colhead{Cross} &
\colhead{Exposure} &
\colhead{Wavelengths} &
\colhead{SNR} &
\colhead{SNR} \\
\colhead{} &
\colhead{Disperser} &
\colhead{Time (s)} &
\colhead{Covered (\AA)} &
\colhead{4620 \AA\tablenotemark{a}} &
\colhead{3460 \AA\tablenotemark{b}}
}

\startdata
Dec. 1995 & Red & 7200 & 3650-6080 & 10 & --\tablenotemark{c} \nl
Dec. 1996 & Red & 6400 & 3650-6080 &  8 & --\tablenotemark{c} \nl
Dec. 1996 & Red & 4300 & 3650-6080 &  5 & --\tablenotemark{c} \nl
Sep. 1998 & UV & 7200 & 3340-4880 & 25 & 8 \nl
Sep. 1998 & UV & 7200 & 3340-4880 & 25 & 9 \nl
Oct. 1998 & UV & 5400 & 3370-4910 & 20 & 6 \nl
Oct. 1998 & UV & 9000 & 3370-4910 & 29 & 10 \nl
Oct. 1998 & UV & 3900 & 3370-4910 & 15 & 4 \nl
Oct. 1998 & UV & 1800 & 3370-4910 & 4 & 1 \nl
Oct. 1998 & UV & 7200 & 3370-4910 & 13 & 4 \nl
Oct. 1998 & UV & 7200 & 3370-4910 & 12 & 3 \nl
Oct. 1998 & UV & 9000 & 3370-4910 & 20 & 6 \nl
Oct. 1998 & UV & 4438 & 3370-4910 & 12 & 4 
\enddata

\tablenotetext{a}{The $z_{abs} \sim 2.8$ \lya\ line falls at 4620 \AA\ where the 
total signal to noise ratio is about 60 per 2.1 \kms.}
\tablenotetext{b}{The $z_{abs} \sim 2.8$ Lyman limit falls at 3460 \AA\ where the 
total signal to noise ratio is about 19 per 2.1 \kms.}
\tablenotetext{c}{The Red- cross-disperser observations of \object\ do not cover
	the $z_{abs} \sim 2.8$ Lyman Limit}
\end{deluxetable}

\begin{deluxetable}{cccc}
\tablecaption{\label{dhlinetab} Ions observed in 
	the $z \sim 2.8$ LLS towards \object}
\tablewidth{34pc}

\tablehead{
\colhead{Ion} &
\colhead{log N} &
\colhead{$b$} &
\colhead{$z$} 
\\
\colhead{} & \colhead{(\cmm)} & \colhead{(\kms)} & \colhead{}
}

\startdata
H I 
& $16.66 \pm 0.02$ & $22.7 \pm 0.1$ & $2.799742 \pm 0.000003$ \nl

C IV                   
& $12.77 \pm 0.08$ & $6.1 \pm 0.9$  & $2.799629 \pm 0.000010$ \nl

Si~III\tablenotemark{a} 
& $12.24 \pm 0.01$ & $7.5 \pm 0.2$  & $2.799736 \pm 0.000002$ \nl

Si~IV                   
& $12.33 \pm 0.03$ & $6.5 \pm 0.7$  & $2.799659 \pm 0.000006$ \nl
\enddata

\tablenotetext{a}{The Si~III line may be a two component blend.}
\end{deluxetable}


\end{document}